\documentclass[twocolumn,showpacs,preprintnumbers,amsmath,amssymb]{revtex4-1}
\usepackage{epsfig}
\usepackage{graphicx}
\usepackage{epsfig}
\usepackage{color}
\usepackage{rotating}
\usepackage{amssymb}
\usepackage{amsmath}
\usepackage{graphicx}

\def\beq{\begin{eqnarray}}
\def\eeq{\end{eqnarray}}

\newcommand{\be}{\begin{equation}}
\newcommand{\ee}{\end{equation}}
\newcommand{\bea}{\begin{eqnarray}}
\newcommand{\eea}{\end{eqnarray}}

\begin{document}

\title{On the von Neumann entropy of a bath linearly coupled to a driven quantum system}

\author{Erik Aurell}
\email{eaurell@kth.se}
\affiliation{
Dept. of Computational Biology and ACCESS Linnaeus Centre and Center for Quantum Materials,
KTH -- Royal Institute of Technology,  AlbaNova University Center, SE-106 91~Stockholm, Sweden\\
Depts. Information and Computer Science and Applied Physics, Aalto University, Espoo, Finland
}
\author{Ralf Eichhorn}
\email{eichhorn@nordita.org}
\affiliation{
Nordita, SE-106 91~Stockholm, Sweden
}

\begin{abstract}
The change of the von Neumann entropy of a set of harmonic oscillators initially in thermal
equilibrium and interacting linearly with an externally driven quantum system is computed by
adapting the Feynman-Vernon influence functional formalism. This quantum entropy
production has the form of the expectation value of three functionals of the forward and backward
paths describing the system history in the Feynman-Vernon theory.
In the classical limit of Kramers-Langevin dynamics (Caldeira-Leggett model) 
these functionals combine to three terms, where the first is
the entropy production functional of stochastic thermodynamics,
the classical work done by the system on  the environment in units of $k_BT$,
the second another functional with no analogue in stochastic thermodynamics, and 
the third is a boundary term.
\end{abstract}


\maketitle

\section{Introduction}
The discovery of fluctuation relations~\cite{ES94,GC95,Jarzynski97}
has transformed classical out-of-equilibrium thermodynamics, giving rise to the new
field of stochastic thermodynamics~\cite{Searles08-review,Jarzynski11-review,seifert12-review}. 
A central idea of the field is that classical termodynamics can be extended to single, usually mesoscopic, systems
where a surrounding medium takes the role of a heat bath~\cite{Sekimoto-book}. 
The fundamental quantity in stochastic thermodynamics is $\delta S_{env}$, the entropy production in the
environment. Mathematically, this quantity can be defined as the log-ratio of transition probabilities in a forward and a 
reversed process~\cite{ChG07},
and fluctuation relations then follow as tautological identities~\cite{Maes99,Gawedzki13}. 

A heat bath is an idealization of a large environment relaxing on a much faster time scale than
the system of interest. Hence, the heat bath is arbitrarily close to thermal equilibrium at all times,
and the entropy production of the system is nothing but the change of the (thermodynamic) entropy of the bath.
By Clausius' formula this gives $\delta S_{env}=\beta \delta Q$ where $\delta Q$ is the energy (heat) transferred
from the system to the bath. Physically, fluctuation relations are non-trivial  
because it is not obvious that the log-ratio way of defining $\delta S_{env}$ is the same as $\beta \delta Q$. 
Indeed, although in retrospect a fairly straightforward fact, for standard physical kinetics 
without memory (master equations, diffusion equations) this
equivalence has only been widely appreciated for a decade and a half~\cite{Jarzynski97b,Kurchan98,LS99}. 

Potential extensions of statistical thermodynamics and fluctuation relations to the quantum domain 
have been extensively investigated, and
reviewed in \cite{Esposito09-review,Campisi11-review}. However, with the exception of the 
Jarzynski equality and Crooks' fluctuation theorem in closed quantum 
systems~\cite{Kurchan-unpub}, the results obtained to date lack the generality and simplicity of
fluctuation relations in classical systems. In open quantum systems 
specific assumptions such as that the dynamics is unital (superoperator preserves the unit 
operator)~\cite{RasteginZyczkowski13},
or that the quantum jump method~\cite{Legio13,HekkingPekola13,HorowitzParrondo13}
or the Lindblad formalism~\cite{ChetriteMallick} can be applied, seem to be needed.
The task is nontrivial since entropy production, as work and heat, 
is not a standard quantum operator. Indeed, according to the original proposal 
for the simplest case of closed quantum system work depends on two
quantum measurements~\cite{Kurchan-unpub}, while in a recent proposal involving only one 
measurement~\cite{Roncaglia14}, a second quantum system is needed to keep track of the work. 

The first goal of this contribution is to show that a natural extension to the quantum domain of
the thermodynamic version of $\delta S_{env}$ can be investigated analytically by adapting the method of Feynman and Vernon~\cite{FeynmanVernon}.
This extension, which we will call $\overline{\delta S_{q}}$, is the change of the von Neumann entropy of a heat bath 
between two measurements on the system. 
In the context of stochastic thermodynamics
a quantity equivalent to $\overline{\delta S_{q}}$ was introduced in~\cite{ELB10}, and more recently investigated
in \cite{PucciEspositoPeliti}.

%In the Feynman-Vernon approach the bath is taken to consist of harmonic oscillators initially in thermal
%equilibrium and linearly coupled to the system. 
%The bath variables can then be integrated out and the propagator
%of the reduced density matrix of the system obtained as a double functional integral over forward and backward
%paths of the system variables only. The matrix element of this propagator with initial and final states of the system 
%gives the transition probability for the system to start from an initial state, obtained after a first measurement,
%and then end up in the final state, the outcome of a second measurement. The first order change the von Neumann
%entropy of the bath between the same two measurements on the system is then obtained in a similar manner.

In the Feynman-Vernon method the bath is taken to consist of harmonic oscillators initially in thermal
equilibrium and linearly coupled to the system, which allows for integrating out the bath.
As a result $\overline{\delta S_{q}}$ can be written as the (quantum) expectation value of three functionals of the system history,
similar but not identical to the real and imaginary actions $S_i$ and $S_r$
in the Feynman-Vernon theory. In the classical limit of a Brownian particle, where 
the system development is described by a Kramers-Langevin equation~\cite{CaldeiraLeggett83a},
these expectation values combine to two averages over the (classical) stochastic process,
which will call $\overline{\delta S_{env}}$ and $\overline{\delta S_{var}}$, and a boundary term, which we will call $\Delta S_{b}$.
$\overline{\delta S_{env}}$ is the average over the process and over a finite time of $\delta S_{env}$, the standard entropy production functional in 
stochastic thermodynamics. By Clausius' formula and for Kramers-Langevin dynamics 
$\delta S_{env}$ equals $\beta( \delta Q_{friction} +\delta Q_{noise})$ where
$\delta Q_{friction}$ and $\delta Q_{noise}$ are the amounts of energy (heat) transmitted from the system to the bath 
by respectively the friction force and the random force.
$\overline{\delta S_{var}}$ is also an average over the same stochastic process,
formally the finite part of the square of the random force, and has no
analogue in standard stochastic thermodynamics.
$\Delta S_{b}$ finally depends only on the (classical) transition probability over a finite time interval, and is hence 
not a functional of the whole (classical) system history. As will be discussed below this 
quantity has very different properties than one expects for (classical) entropy production.  

The paper is organized as follows: in Section~\ref{s:vonNeumann} we define $\overline{\delta S_{q}}$,
equation (\ref{eq:dTrLogRho}), and show how to express it as a quantum expectation value, 
equations~(\ref{eq:dTrLogRho-3}) and~(\ref{eq:dTrLogRho-4}).
In Section~\ref{s:FV} we introduce the Feynman-Vernon formalism and use it to
give an expression for the expectation value, equation (\ref{eq:R-definition}).
In Section~\ref{s:evaluation} we evaluate this expression and in parallel give 
standard results of the Feynman-Vernon theory. The three functionals mentioned above
then appear in equations (\ref{eq:I-definition}) and (\ref{eq:h-definition}).
In Section~\ref{s:CL} we introduce the Caldeira-Leggett limit of the Feynman-Vernon model
which leads to classical dynamics with noise and friction; the limit of the three functionals 
is given in equation (\ref{eq:I-definition-CL}).
In Section~\ref{s:analysis} we analyze the three functionals in this limit
and separate out $\Delta S_{b}$,  
and in Section~\ref{s:interpretation} we 
group the remaining parts into $\overline{\delta S_{env}}$ and $\overline{\delta S_{var}}$.
For completeness we also give, in Section~\ref{s:analysis}, a derivation of the Caldeira-Leggett result that the 
Wigner transform of the Feynman-Vernon propagator for the density matrix goes to the transition 
probability of the (classical) stochastic process.
In Section~\ref{s:asymptotic-analysis} we consider $\Delta S_{b}$ in the limit of weak
coupling between the system and the bath, and in Section~\ref{s:discussion} we sum up and discuss our results.
In Appendices~\ref{s:times} and~\ref{s:higher} we discuss for completeness
the time scales involved and higher-order corrections to the Caldeira-Leggett limit. 

We end this Introduction by noting that in the classical limit we are limited to averages of the entropy production over a finite
time, which only approximates the entropy production functional of stochastic thermodynamics if the time is short.
The question of whether quantum entropy production defined as in this paper can also lead to another definition in
terms of forward and reversed (quantum) dynamics is left for future work. 

\section{The first-order change of von Neumann entropy}
\label{s:vonNeumann}
We consider the setting where a quantum system is prepared in an initial pure state $|i><i|$ at time $t_i$ and then attached
to a bath with density operator $\rho_B^{eq}$ describing a state of thermal equilibrium. Over a time period $[t_i,t_f]$ the system and the bath develop in interaction such that the total
state at time $t_f$ is $\rho_{TOT}^{f}$. At this point a measurement is made of an operator $O$ which depends on the system variables only,
with outcome $o_f$, corresponding to the pure state $|f><f|$ of the system. This happens with probability $P_{if}=\hbox{Tr}_B<f|\rho_{TOT}^f|f>$.
By the measurement postulate the total state after the measurement is 
$\rho_{TOT}^{f,+}=\frac{1}{P_{if}}|f><f|\oplus \left<f|\rho_{TOT}^{f}|f\right>$, and we can therefore identify
the density operator of the bath, after interacting with the system and after the measurement has been performed on the system,
as $\rho_{B}^{f}=\frac{1}{P_{if}}\left<f|\rho_{TOT}^{f}|f\right>$. 
We assume $\delta \rho_B = \rho_B^f - \rho_B^{eq}$ to be small, and
the first-order change of the bath entropy is then
\begin{equation}
\label{eq:dTrLogRho}
\delta\left(\hbox{Tr}_B[-\rho_B \log\rho_B]\right)=\hbox{Tr}_B[-\delta\rho_B \log \rho_B^{eq}]
\end{equation}
%where the first term on the right-hand side can be written
%\begin{equation}
%\label{eq:dTrLogRho-2}
%-\frac{\hbox{Tr}_B[<f|\rho_{TOT}^{f}|f> \log \rho_B^{eq}]}{P_{if}}+\hbox{Tr}_B[\rho_B^{eq} \log \rho_B^{eq}]
%\end{equation}
%The second term in (\ref{eq:dTrLogRho-2}) does not depend on the system, and is simply (minus) the standard 
%thermodynamic entropy of the bath, $\sum_i -S_i$.
Equation (\ref{eq:dTrLogRho}) is our definition of $\overline{\delta S_q}$.
In a basis of energy eigenstates $|\underbar{n}>=|n_1,n_2,\ldots>$ the density operator 
$\rho_B^{eq}$ is diagonal with elements
$<\underbar{n}|\rho_B^{eq}|\underbar{n}>=\prod_b e^{-\beta \left(E^b_{n_b}-F_b(\beta)\right)}$
where $E^b(n_b)$ is the energy of the $b$’th degree of freedom of the bath in state $n_b$, and
$F_b$ is its free energy at inverse temperature $\beta$.
The expression in (\ref{eq:dTrLogRho}) can therefore be written
\begin{equation}
\overline{\delta S_q}=\beta\frac{d}{d\epsilon}\log R_{if}(\epsilon)|_{\epsilon=0}-\beta\sum_b U_b(\beta)
\label{eq:dTrLogRho-3}
\end{equation}
%\begin{equation}
%-\frac{1}{P_{if}}\frac{\beta d}{d\epsilon}
%\left(\sum_{\underbar{n},\underbar{m}}<\underbar{n},f|\rho_{TOT}^{f}|f,\underbar{m}><\underbar{m}|e^{-\epsilon H_B}|\underbar{n}>
%\right)_{\epsilon=0} -\beta\sum_b U_b(\beta)
%\label{eq:dTrLogRho-3}
%\end{equation}
where 
%and the two last terms combine to give $-\sum_i \beta U_i(\beta)$,
$U_b(\beta)=\frac{\omega_b\hbar}{2}\coth(\frac{\omega_b\beta\hbar}{2})$ is the internal energy of oscillator $b$,
%The first term on the right-hand side of (\ref{eq:dTrLogRho-3}) 
%can be expressed as $\beta\frac{d}{d\epsilon}\log R_{if}(\epsilon)|_{\epsilon=0}$ where
\begin{equation}
\label{eq:dTrLogRho-4}
R_{if}(\epsilon) = \hbox{Tr}_B[\left<f|\rho_{TOT}^f|f\right>e^{-\epsilon H_B}],
\end{equation}
and $H_B$ is the Hamiltonian of the bath.
Note that $R_{if}(0)=P_{if}$. 
Equations (\ref{eq:dTrLogRho-3}) and (\ref{eq:dTrLogRho-4}) are the starting point of our analysis.

\section{The Feynman-Vernon method}
\label{s:FV}
The Feynman-Vernon theory is a means to compute $P_{if}$ while
we need to compute the slightly more complex quantity $R_{if}'(0)$.
Let us begin by noting that the time development of an open quantum system is described by a superoperator 
or quantum map $\Phi$ which maps density operators to density operators, and which
can always be realized by adding another system or ancilla in state $\rho_a$, acting unitarily on the
combined system and ancilla, and then tracing out the ancilla 
$\Phi\rho=\hbox{Tr}_a[U\left(\rho\oplus\rho_a \right)U^{\dagger}]$~\cite{BreuerPetruccione,BengtssonZyczkowski,ChruscinskiSarbicki14}.
The Feynman-Vernon approach consist in writing the two unitary operators
$U$ and $U^{\dagger}$ as path integrals while taking the ancilla to be a 
bath of harmonic oscillators initially in thermal equilibrium, linearly coupled to the system.
The total Hamiltonian describing the system and ancilla is thus
\begin{equation}
H=H_S(X,P,t) - X\sum_b C_b x_b+ \sum_b \frac{p_b^2}{2m_b}+\frac{m_b\omega_b^2}{2}x_b^2
\label{eq:Feynman-Vernon-H}
\end{equation}
where $C_b$ is the strength of the interaction between the system and bath oscillator $b$. 
Integrating out the bath gives $\Phi$ in the coordinate representation as
\bea
&&[\Phi\rho(t_i)](X^f,Y^f)=\rho(X^f,Y^f,t_f)=  \nonumber \\
&&\int dX^idY^i K(X^f,Y^f,t_f;X^i,Y^i,t_i)\rho(X^i,Y^i,t_i)
\eea
and the transition probability as
\begin{eqnarray}
P_{if}&=& \int dX^idY^i dX^f dY^f \psi_i(X^i)\psi_i^*(Y^i)\nonumber \\
&&\psi_f^*(X^f)\psi_f(Y^f)K(X^f,Y^f,t_f;X^i,Y^i,t_i)
\end{eqnarray}
where $\psi_i$ and $\psi_f$ are the initial and final states in the coordinate representation and
$K$ is the Feynman-Vernon propagator of the density operator of the system~\cite{FeynmanVernon}.
The first step in computing $K$ is the path integrals over the bath in
$U\left(\rho\oplus\rho_a \right)U^{\dagger}$ with fixed initial and final position of each
bath oscillator. The result of these path integrals is
$K_b(q_f,t_f,q_i,t_i;X) K_b^*(q_f',t_f,q_i',t_i;Y)$ where the $K_b$'s are
propagators of harmonic oscillators with linear terms in the action, 
$\int X(s) C_b x_b(s)$ for the forward path, 
and $\int Y(s) C_b y_b(s)$ for the backward path.
The second step is to introduce a coordinate representation of     
the equilibrium density operator of the bath oscillator, proportional to the propagator in imaginary time.
Integrating out the four positions $q_i$, $q_f$, $q_f'$ and $q_i'$ then gives
the Feynman-Vernon influence functional ${\cal F}_b[\{X(s)\},\{Y(s)\}]$ as a four-dimensional integral
\begin{eqnarray}
{\cal F}_b&=&\int d\underline{q}\frac{1}{Z_b(\beta)}K_b(q_i,0,q_i',i\beta\hbar;0)\delta(q_f-q_f')
\nonumber \\
&&K_b(q_f,t_f,q_i,t_i;X) K_b^*(q_f',t_f,q_i',t_i;Y)
\label{eq:influence}
\end{eqnarray}
where $Z_b=\left(2\sinh\left(\frac{\omega_b\epsilon\hbar}{2}\right)\right)^{-1}$
is the partition function of $b$ at inverse temperature $\beta$, and
the delta function is the coordinate representation of the trace over the final state of $b$.
The Feynman-Vernon propagator can then be written as double path integral over
the system variables only
\begin{equation}
K=\int {\cal D}X{\cal D}Ye^{\frac{i}{\hbar}\left(S_S[X]-S_S[Y]\right)}\prod_b{\cal F}_b
\label{eq:Feynman-Vernon-K}
\end{equation}
A comparison of (\ref{eq:dTrLogRho-4}) and (\ref{eq:influence}) shows that to compute the 
first-order change of the von Neumann entropy of the bath we need
\begin{eqnarray}
{\cal F}_b^{\epsilon}&=&\int d\underline{q}\frac{1}{Z_b(\beta)}K_b(q_i,0,q_i',i\beta\hbar;0)K_b(q_f,0,q_f',i\epsilon\hbar;0)
\nonumber \\
&& K_b(q_f,t_f,q_i,t_i;X) K_b^*(q_f',t_f,q_i',t_i;Y)
\label{eq:influence-2}
\end{eqnarray}
in terms of which
\begin{eqnarray}
R_{if}'(0)&=& \int dX^idY^i dX^f dY^f \psi_i(X^i)\psi_i^*(Y^i)\psi_f^*(X^f)\psi_f(Y^f) 
\nonumber \\
&{\cal D}X&{\cal D}Ye^{\frac{i}{\hbar}\left(S_S[X]-S_S[Y]\right)}\prod_b{\cal F}_b
\frac{d}{d\epsilon}\left(\sum_b\log{\cal F}_b^{\epsilon}\right)_{\epsilon=0} 
\label{eq:R-definition}
\end{eqnarray}

\section{Evaluating the entropy production functional}
\label{s:evaluation}
The propagator of a harmonic oscillator is the exponential of terms 
constant, linear and quadratic in the 
initial and final position.
After integrating out these initial and final positions the
modified influence functional is 
\begin{equation}
{\cal F}_b^{\epsilon}=
\frac{(\frac{2\pi\hbar}{\omega m})^2e^{-\frac{1}{2m\omega\hbar}\underline{u}M^{-1}(\epsilon)\underline{u}+\frac{i}{\hbar}B}}
{Z_b(\beta)N_b(i\beta\hbar) |N_b(t)|^2 N(i\epsilon\hbar)(\det M(\epsilon))^{\frac{1}{2}}}
\label{eq:influence-3}
\end{equation}
where  $t=t_f-t_i$, 
$N_b(t)=\sqrt{\frac{2\pi i\hbar\sin(\omega t)}{m \omega }}$
are the normalizations of the propagators, 
the constant term $B$ is given by 
%\begin{widetext}
\begin{eqnarray}
&-&\frac{C^2}{m\omega\sin\omega t}\int^{t_f}\!\!\int^s\! ds ds'
\sin\omega (t_f-s)\sin\omega s' X(s)X(s')\nonumber \\
 &+&\frac{C^2}{m\omega \sin\omega t}\int^{t_f}\!\!\int^s\! ds ds'
\sin\omega (t_f-s)\sin\omega s' Y(s)Y(s')\nonumber
\end{eqnarray}
%\end{widetext}
and the matrix $M(\epsilon)$ and the vector $\underline{u}$ are
respectively given by
\begin{widetext}
\begin{equation}
\left(\begin{array}{cccc}
\frac{1}{i}\cot\omega t+\coth\omega\beta\hbar & -\sinh^{-1}\omega\beta\hbar & i\sin^{-1}\omega t & 0 \\ 
-\sinh^{-1}\omega\beta\hbar & i\cot\omega t+\cosh\omega\beta\hbar & 0 & -i\sin^{-1}\omega t \\ 
i\sin^{-1}\omega t  & 0 & \frac{1}{i}\cot\omega t+\coth \omega\epsilon\hbar & -\sinh^{-1}\omega\epsilon\hbar \\ 
0 & -i\sin^{-1}\omega t & -\sinh^{-1} \omega\epsilon\hbar & i\cot\omega t +\coth \omega\epsilon\hbar
\end{array}
\right) \hbox{and}
\left(\begin{array}{l}
  \frac{C }{\sin\omega t}\int^{t_f}\!ds\sin\omega (t_f-s)X(s)\\
  \frac{-C }{\sin\omega t}\int^{t_f}\!ds\sin\omega (t_f-s)Y(s)\\
  \frac{C }{\sin\omega t}\int^{t_f}\!ds\sin\omega sX(s)\\
  \frac{-C }{\sin\omega t}\int^{t_f}\!ds\sin\omega sY(s)
\end{array}\right).
\nonumber
\end{equation}
\end{widetext}
%and
%\begin{equation}
%\underline{u}
%=\left(\begin{array}{l}u_i\\u_i'\\u_f\\u_f'\end{array}\right)
%\left(\begin{array}{l}
%  \frac{C }{\sin\omega t}\int^{t_f}ds\sin\omega (t_f-s)X(s)\\
%  \frac{-C }{\sin\omega t}\int^{t_f}ds\sin\omega (t_f-s)Y(s)\\
%  \frac{C }{\sin\omega t}\int^{t_f}ds\sin\omega sX(s)\\
%  \frac{-C }{\sin\omega t}\int^{t_f}ds\sin\omega sY(s)
%\end{array}\right)
%\nonumber
%\end{equation}
In above $B$, $C$, $\omega$ and $m$ are as for oscillator $b$.
The pre-factors of the exponential in
(\ref{eq:influence-3}) combine to $\sinh(\frac{\omega\hbar\beta}{2})\sinh^{-1}(\frac{\omega\hbar(\beta+\epsilon)}{2})$
which contributes to the derivative $\frac{d}{d\epsilon}{\cal F}_b^{\epsilon}|_{\epsilon=0}$ 
a term $-\frac{\omega\hbar}{2}\coth\frac{\omega\hbar\beta}{2}$.
This cancels with the
term $U_b$ in (\ref{eq:dTrLogRho-3}), and the quantity sought is thus
\begin{widetext}
\begin{equation}
\label{eq:dTrLogRho-5}
\overline{\delta S_{q}}=\frac{1}{P_{if}}\int dX^idY^i dX^f dY^f \psi_i(X^i)\psi_i^*(Y^i)\psi_f^*(X^f)\psi_f(Y^f)
{\cal D}X{\cal D}Ye^{\frac{i}{\hbar}S_S[X]-\frac{i}{\hbar}S_S[Y]}\prod_b{\cal F}_b
\sum_b\left(\frac{-\beta}{2m_b\omega_b\hbar}\underline{c_b}M_b^{-1\prime}(0)\underline{c_b}\right)
\end{equation}
\end{widetext}
Equation (\ref{eq:dTrLogRho-5}) is the first result of this paper giving the
entropy production in the bath between two measurements on the system  
as the expected value of a functional of the forward and backward
system paths, divided by the transition probability. 
To proceed further we note that the product $\prod_b{\cal F}_b$, which is 
$e^{ \sum_b\left(-\frac{1}{2m_b\omega_b\hbar}\underline{c_b}M_b^{-1}(0)\underline{c_b}+\frac{i}{\hbar}B_b\right)}$,
can be written $e^{\frac{i}{\hbar}S_i-\frac{1}{\hbar}S_r}$ defining the real 
and imaginary parts of the Feynman-Vernon influence action~\cite{FeynmanVernon,CaldeiraLeggett83a}.
These are quadratic functionals of the paths of the system,
%\begin{widetext}
%\begin{eqnarray}
%\label{eq:S_i-defintion}
%S_i[X,Y] = \int^{t_f}\int^{\tau}\left(X(\tau)-Y(\tau)\right)\left(X(s)+Y(s)\right)k_i(\tau-s)&\quad& 
%       k_i=\sum_b\frac{C_b^2}{2m_b\omega_b}\sin\omega_b(\tau-s) \\
%\label{eq:S_r-defintion}
%S_r[X,Y] = \int^{t_f}\int^{\tau}\left(X(\tau)-Y(\tau)\right)\left(X(s)-Y(s)\right)k_r(\tau-s)&\quad&
%k_r=\sum_b\frac{C_b^2}{2m_b\omega_b}\coth(\frac{\beta\hbar\omega_b}{2})\cos\omega_b(\tau-s)
%\end{eqnarray}
%\end{widetext}
\begin{eqnarray}
\label{eq:S_i-defintion}
S_i&=& \int^{t_f}\!\!\int^{s}\!(X-Y)(X'+Y')k_i(s-s')ds' ds\nonumber \\
\label{eq:S_r-defintion}
S_r&=& \int^{t_f}\!\!\int^{s}\!(X-Y)(X'-Y')k_r(s-s')ds' ds\nonumber
\end{eqnarray}
where we write $X$ and $Y$ for quantities at time $s$ and
$X'$ and $Y'$ for quantities at the earlier time $s'$, and where
the kernels are
\begin{eqnarray}
\label{eq:k_i-defintion}
k_i(s-s')&=&\sum_b\frac{C_b^2}{2m_b\omega_b}\sin\omega_b(s-s')\nonumber \\
\label{eq:k_r-defintion}
k_r(s-s')&=&\sum_b\frac{C_b^2}{2m_b\omega_b}\coth(\frac{\beta\hbar\omega_b}{2})\cos\omega_b(s-s') \nonumber
\end{eqnarray}
By a similar analysis as the one leading to $S_i$ and $S_r$ the functional 
in (\ref{eq:dTrLogRho-5})
can be evaluated to ${\cal I}^{(1)}[X,Y]+{\cal I}^{(2)}[X,Y]+{\cal I}^{(3)}[X,Y]$ where
%\begin{widetext}
%\begin{equation}
%\begin{array}{lll}
%{\cal I}^{(1)} &= \int^{t_f} \int^{s} (X-Y)(X'-Y')h^{(1)}(s-s') \quad 
%                &h^{(1)}=-\sum_b\frac{\beta C_b^2}{4m_b}\sinh^{-2}(\frac{\beta\hbar\omega_b}{2})\cos\omega_b(s-s')\\ 
%{\cal I}^{(2)} &= \int^{t_f} \int^{s} (XY'-X'Y) h^{(2)}(s-s') \quad 
%               &h^{(2)}= i\sum_b\frac{C_b^2}{2m_b}\coth(\frac{\beta\hbar\omega_b}{2})\sin\omega_b(s-s') \\
%{\cal I}^{(3)} &= \int^{t_f} \int^{s}  (XY'+X'Y) h^{(3)}(s-s')\quad
%                &h^{(3)}=\sum_b\frac{\beta C_b^2}{2m_b}\cos\omega_b(s-s')\\
%\end{array}
%\label{eq:I-definition}
%\end{equation}
%\end{widetext}

\begin{equation}
\begin{array}{ll}
{\cal I}^{(1)} &= \int^{t_f} \int^{s} (X-Y)(X'-Y')h^{(1)}(s-s')ds' ds \\ 
{\cal I}^{(2)} &= \int^{t_f} \int^{s} (XY'-X'Y) h^{(2)}(s-s')ds' ds \\
{\cal I}^{(3)} &= \int^{t_f} \int^{s}  (XY'+X'Y) h^{(3)}(s-s')ds' ds 
\end{array}
\label{eq:I-definition}
\end{equation}
with the kernels
\begin{equation}
\begin{array}{ll}
h^{(1)}&=-\sum_b\frac{\beta C_b^2}{4m_b}\sinh^{-2}(\frac{\beta\hbar\omega_b}{2})\cos\omega_b(s-s')\\ 
h^{(2)}&= i\sum_b\frac{\beta C_b^2}{2m_b}\coth(\frac{\beta\hbar\omega_b}{2})\sin\omega_b(s-s') \\
h^{(3)}&=\sum_b\frac{\beta C_b^2}{2m_b}\cos\omega_b(s-s')
\end{array}
\label{eq:h-definition}
\end{equation}
Equations (\ref{eq:I-definition}) and (\ref{eq:h-definition})     
is the second result of this paper. We note 
that while 
${\cal I}^{(1)}$ equals $-\frac{1}{\hbar}\beta\partial_{\beta}S_r$,
${\cal I}^{(2)}$ and ${\cal I}^{(3)}$ 
are new and non-causal terms,
\textit{i.e.} which do not fullfill General Property 5 of 
influence functionals as discussed on pp 126-127 in \cite{FeynmanVernon}. 
For a physical interpretation
we turn to the limit of classical stochastic dynamics.

\section{The Caldeira-Leggett limit}
\label{s:CL}
The classical limit of the Feynman-Vernon theory was computed in~\cite{CaldeiraLeggett83a}.
The spectrum of the bath oscillators is then assumed continuous with density $f(\omega)$ 
such that $f(\omega)C^2_{\omega}/m(\omega)$ equals $\frac{2\eta\omega^2}{\pi}$ 
up to some upper cut-off $\Omega$.
The parameter $\eta$ has the dimension $ML/T$ of a classical friction coefficient,
and the first kernel in the Feynman-Vernon theory then tends to
$k_i\approx - \eta\frac{d}{d(s-s')}\delta(s-s')$.
The corresponding action $S_i$ is a potential renormalization 
plus a term $-\frac{\eta}{2}\int (X-Y)(\dot{X}+\dot{Y})ds$.
As a stochastic integral this finite part of $S_i$ has to be interpreted in the post-point (anti-It\^o)
prescription since the time derivatives stem from the integral over $s'$ up to $s$. 
The other kernel $k_r$ 
%$k_r=\frac{\eta}{\pi}\int^{\Omega} d\omega \omega \coth(\frac{\beta\hbar\omega}{2})\cos(\omega(\tau-s))$
has only small contributions for $|\omega|>\frac{1}{\beta\hbar}$ and hence describes, if $\Omega$ is large enough, 
a memory kernel of width $\beta\hbar$, independently of $\Omega$. 
If all times of interest in the system are longer than $\beta\hbar$ then
$S_r\approx \frac{\eta}{\beta\hbar}\int (X-Y)^2ds$.
The different time scales involved are briefly discussed in Appendix~\ref{s:times}
and possible higher-order corrections in Appendix~\ref{s:higher}.

In the same limit as above the three terms in (\ref{eq:I-definition})
tend to
\begin{equation}
\begin{array}{lrl}
{\cal I}^{(1)} &=&-\frac{\eta}{\beta\hbar^2} \int^{t_f} (X-Y)^2\, ds=-\frac{1}{\hbar}S_r \\
{\cal I}^{(2)} &=&i\frac{\eta}{\hbar}        \int^{t_f} (\dot{X}Y-X\dot{Y})\,ds \\
{\cal I}^{(3)} &=&\beta\eta                  \int^{t_f} \dot{X}\dot{Y}ds
\end{array}
\label{eq:I-definition-CL}
\end{equation}
The integral for ${\cal I}^{(2)}$ in (\ref{eq:I-definition}) can be extended over the whole domain
-- \textit{i.e.} $\int^{t_f}\int^{t_f}\left(\cdots\right)ds' ds$ -- and expanded around the diagonal ($s=s'$).
Contrary to the case of $S_i$ there is therefore not any potential renormalization term
from  ${\cal I}^{(2)}$ in the Caldeira-Leggett limit.
Furthermore, the limit of  ${\cal I}^{(2)}$ given in (\ref{eq:I-definition-CL}) does not
depend on the discretization
scheme since the It\^o contributions cancel. Therefore, we can alternatively write this term as
\begin{equation}
{\cal I}^{(2)} = \frac{2i\eta}{\hbar}S_i^{(mid)} + \Delta S_b  
\label{eq:S_b-definition-1}
\end{equation}
where
\begin{equation}
\frac{2i\eta}{\hbar}S_i^{(mid)} = -\frac{i\eta}{\hbar}\int (X-Y)(\dot{X}+\dot{Y})ds 
\label{eq:S_i-CL-limit}
\end{equation}
and
\begin{equation}
\Delta S_b =  \frac{i\eta}{\hbar}\int (X\dot{X}-Y\dot{Y})ds  
\label{eq:S_b-definition-2}
\end{equation}
and with the mid-point prescription for both terms. Consquently, the term $\Delta S_b$ 
is to be interpreted as
\begin{equation}
\Delta S_b = \frac{i\eta}{2\hbar}\left[ X^2-Y^2\right]_i^f 
\label{eq:S_b-definition-3}
\end{equation}
which is a pure boundary contribution.

Of the three terms in (\ref{eq:I-definition-CL})
it is the average of ${\cal I}^{(3)}$ which has the most physical immediate 
meaning, as it must tend to $\overline{\beta\eta \int v^2 dt}$ when
$\dot{X}$ and $\dot{Y}$ are approximated by a classical velocity $v$.
To a friction force $-\eta v$ corresponds a reaction force $\eta v$ from the system on the bath,
and the work done by this force is $\eta \int v^2 dt$. The entropy production from  ${\cal I}^{(3)}$ 
is therefore $\overline{\beta\delta Q_{friction}}$, where $\delta Q_{friction}$ is the energy (heat) transferred from the system
to the bath, as announced in Introduction. 

\section{Analysis of the classical limit}
\label{s:analysis}
To further analyze the classical limit we use, as in~\cite{CaldeiraLeggett83a},
the Markov property of Feynman-Vernon propagator, in this limit.
First it is convenient to introduce auxiliary variables $q=\frac{X+Y}{2}$,
$\alpha=Y-X$ and $\tilde K(q_f,\alpha_f,q_i,\alpha_i)=K(X_f,Y_f,X_i,Y_i)$,
and the Wigner transform of $K$, written as
\begin{equation} 
P=\left(\frac{1}{2\pi\hbar}\right)^d\int e^{\frac{ip_f\alpha_f}{\hbar}-\frac{ip_i\alpha_i}{\hbar}}
\tilde K(q_f,\alpha_f,q_i,\alpha_i,t_f,t_i)
\label{eq:Wigner-2}
\end{equation}
This $P$ satisfies the Fokker-Planck equation, and is hence the transition probability of a classic
stochastic process starting at $(q_i,p_i)$ at time $t_i$, and ending up in $(q_f,p_f)$ at time 
$t_f$~\cite{CaldeiraLeggett83a} (see below). 
It is further convenient to introduce the two overlaps
\begin{eqnarray}
\label{eq:f_i-definition}
f_i(q_i,p_i) &=& \int\! d\alpha_i e^{\frac{ip_i\alpha_i}{\hbar}}\psi_i(q_i\!-\!\frac{\alpha_i}{2}) \psi^*_i(q_i\!+\!\frac{\alpha_i}{2})  \\ 
\label{eq:f_f-definition}
f_f(q_f,p_f) &=& \int\! d\alpha_f e^{\frac{ip_f\alpha_f}{\hbar}}\psi_f(q_f\!-\!\frac{\alpha_f}{2}) \psi^*_f(q_f\!+\!\frac{\alpha_f}{2}) 
\end{eqnarray}
which we assume to be as for coherent states integrated against functions
depending sufficiently weakly on phase space, \textit{i.e.}
$f_i\approx \left(2\pi\hbar\right)^d\delta(q_i-\overline{q_i}) \delta(p_i-\overline{p_i})$
and $f_f\approx \left(2\pi\hbar\right)^d\delta(q_f-\overline{q_f}) \delta(p_f-\overline{p_f})$.
The quantum mechanical transition probability $P_{if}$ is then approximately
$2\pi\hbar P(\overline{q_f},\overline{p_f},\overline{q_i},\overline{p_i},t_f,t_i)$.

For a system with Hamiltonian $H_S=\frac{P^2}{2M}+V(X)$
the short-time density matrix propagator, for $\tilde K$ and from $(q,\alpha)$ at time 
$t_i=t_f-\Delta  s$ to $(q',\alpha')$ at time $t_f$, is    
\begin{eqnarray} 
\tilde K_{\Delta s}(\cdot)&=&\left(\frac{1}{2\pi\hbar\Delta s/M}\right)^d
e^{-\frac{iM\Delta q\Delta\alpha}{\hbar\Delta s}+\frac{i}{\hbar}\eta\alpha'\Delta q}\nonumber \\
&& e^{\Delta s\left(-\frac{i}{\hbar}V(q'-\frac{\alpha'}{2})
                         +\frac{i}{\hbar}V(q'+\frac{\alpha'}{2})-
                           \frac{\eta}{\beta\hbar^2}(\alpha')^2\right)}
\label{eq:short-time}
\end{eqnarray}
where $\Delta q=q'-q$ and  $\Delta \alpha=\alpha'-\alpha$,
the arguments on the left hand side are understood
and the contribution from $S_i$ has been evaluated with the post-point prescription.
The Chapman-Kolmogorov equation for $\tilde K$ is, expanding in the increments, 
\begin{eqnarray} 
\tilde K(\cdot)&=&\int d\Delta q d\Delta\alpha 
\tilde K_{\Delta s}(q_f,\alpha_f,q_f-\Delta q,\alpha_f-\Delta\alpha,\cdot) \nonumber \\
&& \sum_{nm}\frac{(-\Delta q)^n(-\Delta\alpha)^m}{n!m!}\partial_{q}^{n}\partial_{\alpha}^{m}  
\tilde K(\cdot,t_f-\Delta s,t_i)
\label{eq:expansion}
\end{eqnarray}
The integral over $\Delta q$ of a term proportional to $(\Delta q)^n$ 
in (\ref{eq:expansion})
gives, using (\ref{eq:short-time}),
$(\frac{i\hbar\Delta s}{M})^n\delta^{(n)}(\Delta\alpha-\frac{\eta}{M}\alpha'\Delta s)$.
A term proportional to $(\Delta \alpha)^m$ in (\ref{eq:expansion})
can be expanded as  
$\sum_l \left(m\atop l\right)
(\Delta\alpha-\frac{\eta}{M}\alpha'\Delta s)^l(\frac{\eta}{M}\alpha'\Delta s)^{m-l}$
and when integrated over $\Delta\alpha$ this gives zero unless $l=n$.
Using $m=l=0$, $m=1$ and $l=0$ and $m=l=1$ 
equation (\ref{eq:expansion}) can hence be written
\begin{eqnarray} 
\partial_{t}\tilde K &=& \left(-\frac{i}{\hbar}V(q-\frac{\alpha}{2})
                         +\frac{i}{\hbar}V(q+\frac{\alpha}{2})-
                           \frac{\eta}{\beta\hbar^2}(\alpha)^2\right)\tilde K \nonumber \\
&& -i\frac{\hbar}{M}\partial^2_{q\alpha}\tilde K - \frac{\eta}{M}\alpha\partial_{\alpha}\tilde K
\label{eq:CL-Lindbladeq}
\end{eqnarray}
which is the Lindblad equation derived in~\cite{CaldeiraLeggett83a}.
For the Wigner transform (\ref{eq:CL-Lindbladeq}) gives
\begin{eqnarray}
\partial_t P&=& -\partial_q(\frac{p}{M}P)+ \eta\partial_p\left(\frac{p}{M}P\right)+\frac{\eta}{\beta}\partial_{pp}P \nonumber \\
&&+\partial_p(V'(q)P)+{\cal O}(\hbar^2,V^{'''})
\label{eq:CL-FokkerPlanckeq}
\end{eqnarray}
which is, up to terms of order $\hbar^2$, the Fokker-Planck equation of 
classical stochastic dynamics with friction coefficient $\eta$~\cite{CaldeiraLeggett83a}. 
We proceed to treat the three functionals in (\ref{eq:I-definition-CL}) in
an analogous manner. Higher-order corrections to Fokker-Planck equation derived here
(finite $\hbar\beta$ corrections) are briefly discussed in Appendix~\ref{s:higher}.

\subsection{The ${\cal I}^{(1)}$ contribution}
\label{ss:I1-contribution}
At given initial and final positions of the system,
$(q_i,\alpha_i)$ and $(q_f,\alpha_f)$, the path integrals 
in (\ref{eq:dTrLogRho-5}) give for the ${\cal I}^{(1)}$ part
\begin{eqnarray}
&&\int ds \int dq d\alpha \tilde K(q_f,\alpha_f,q,\alpha,t_f,s)(-\frac{\eta}{\beta\hbar^2} \alpha^2) \nonumber \\
&& \tilde K(q,\alpha,q_i,\alpha_i,s,t_i) \nonumber
\end{eqnarray}
Combining this with the integrals over initial and final positions
and (\ref{eq:f_i-definition}) and (\ref{eq:f_f-definition}) we can write 
\begin{eqnarray}
&&\int dq_i dp_i dq_f dp_f f_i f_f^* \int ds \int dq d\alpha dp dp' P(q_f,p_f,q,p',t_f,s)\nonumber \\
&&\frac{1}{(2\pi\hbar)^{2d}}(-\frac{\eta}{\beta\hbar^2} \alpha^2) e^{\frac{ip'\alpha}{\hbar}-\frac{ip\alpha}{\hbar}}
P(q,p,q_i,p_i,s,t_i) \nonumber
\end{eqnarray}
The term $(-\frac{1}{\hbar^2} \alpha^2)$ can be interpreted as $\partial^2_{p'p'}$ acting on $e^{\frac{ip'\alpha}{\hbar}}$
and by integration by parts this gives
\begin{eqnarray}
&&\frac{1}{(2\pi\hbar)^d} \int dq_i dp_i dq_f dp_f f_i f_f^* \int ds \int dq dp  \nonumber \\
&&\frac{\eta}{\beta}\partial^2_{pp}P(q_f,p_f,q,p,\cdot) P(q,p,q_i,p_i,\cdot) \nonumber
\end{eqnarray}
The classical limit of the contribution of $\overline{\delta S_{q}}$  from ${\cal I}^{(1)}$ is therefore
\begin{equation}
\frac{\int ds \int dq dp P(q,p,\overline{q_i},\overline{p_i},s,t_i)\frac{\eta}{\beta}\partial^2_{pp}P(\overline{q_f},\overline{p_f},q,p,t_f,s)}
{P(\overline{q_f},\overline{p_f},\overline{q_i},\overline{p_i},t_f,t_i)}
\label{eq:classical-I1}
\end{equation}
In Section~\ref{s:interpretation} below we show how this can be given an interpretation as
an average $\overline{\delta S_{var}}$ over the stochastic process. 

\subsection{The $\frac{2i}{\hbar}S_i^{(mid)}$ contribution}
\label{ss:2ioverhbarS_i-contribution}
For the ${\cal I}^{(2)}$ contribution we have from the $\frac{2i}{\hbar}S_i^{(mid)}$ 
part in equation (\ref{eq:S_i-CL-limit})
\begin{eqnarray}
&&\int ds \int dq d\alpha dq' d\alpha' \tilde K(q_f,\alpha_f,q',\alpha',t_f,s)
\frac{2\eta i}{\hbar\Delta s}\Delta q(\alpha \nonumber \\ 
&&+ \frac{\Delta\alpha}{2}) \tilde K_{\Delta s}(q',\alpha',q,\alpha,\cdot) \tilde K(q,\alpha,q_i,\alpha_i,s,t_i) \nonumber
\end{eqnarray}
where we have used the mid-point prescription, following the discussion around equation (\ref{eq:S_b-definition-2}).
Focusing first on the pre-point term (the term in the inner parenthesis proportional to $\alpha$), 
we use, similarly to (\ref{eq:expansion}), the short-time expression (\ref{eq:short-time}) and the expansion
\begin{eqnarray} 
\tilde K(q_f,\alpha_f,q',\alpha',t_f,t) &&= \nonumber \\
\sum_{nm}\frac{(\Delta q)^n(\Delta\alpha)^m}{n!m!}&&\partial_{q}^{n}\partial_{\alpha}^{m} \tilde K(q_f,\alpha_f,q,\alpha,t_f,t)
%\label{eq:expansion-2}
\end{eqnarray}
The integral over $\Delta q$ gives $(\frac{i\hbar\Delta s}{M})^{n+1}\delta^{(n+1)}(\Delta\alpha-\frac{\eta}{M}\alpha'\Delta s)$
for a term in the expansion proportional to $(\Delta q)^n$. 
The only contribution of order $\Delta s$ is $n=0$ and $m=l=1$ which gives 
\begin{eqnarray}
&&\frac{2\eta}{M}\int ds \int dq d\alpha\,  \partial_{\alpha} \tilde K(q_f,\alpha_f,q,\alpha,t_f,s)\nonumber \\  
&& \alpha \tilde K(q,\alpha,q_i,\alpha_i,s,t_i) \nonumber
\end{eqnarray}
Combining this with the integrals over initial and final positions as above
we have 
\begin{eqnarray}
&&\int dq_i dp_i dq_f dp_f f_i f_f^* \int ds \int dq d\alpha\, \partial_{\alpha}\left(
\int dp' e^{\frac{ip'\alpha}{\hbar}} P(\cdot)\right)\nonumber \\
&&\,\frac{1}{(2\pi\hbar)^{2d}}(\frac{2\eta}{M}\alpha) 
\int dp e^{-\frac{ip\alpha}{\hbar}}P(q,p,q_i,p_i,s,t_i)) \nonumber
\end{eqnarray}
The factor $\alpha$ can be interpreted as $-\frac{\hbar}{i}\partial_{p}$ acting on
$e^{\frac{ip\alpha}{\hbar}}$
and the derivative
moved then to the last $P$, while the derivative $\partial_{\alpha}$ brings down $ip'/\hbar$
multiplying the first $P$. Combining these terms gives
\begin{eqnarray}
&&\left(\frac{1}{2\pi\hbar}\right)^{d}\int dq_i dp_i dq_f dp_f f_i f_f^* \int ds \int dq  dp \nonumber \\
&&\partial_{p}P(q_f,p_f,q,p,\cdot)\frac{2\eta }{M}p\partial_{p}P(q,p,q_i,p_i,\cdot)
\end{eqnarray}
The classical contribution from the first term in $\frac{2i}{\hbar}S_i^{(mid)}$ to $\overline{\delta S_{q}}$ is therefore
\begin{eqnarray}
&&\frac{\int\! ds dq dp  P(\overline{q_f},\overline{p_f},q,p,\cdot) \frac{2\eta}{M}p\partial_{p}  
P(q,p,\overline{q_i},\overline{p_i},\cdot)}
{P(\overline{q_f},\overline{p_f},\overline{q_i},\overline{p_i},t_f,t_i)}
\label{eq:classical-Si}
\end{eqnarray}
The second term in (\ref{eq:S_i-CL-limit}) (term in inner parenthesis proportional to $\Delta\alpha$)
is on the other hand
\begin{eqnarray}
&&\int ds \int dq d\alpha dq' d\alpha' \tilde K(q_f,\alpha_f,q',\alpha',t_f,s)
(\frac{\eta i}{\hbar\Delta s})\nonumber \\ 
&&(\Delta\alpha\Delta q) \tilde K_{\Delta s}(q',\alpha',q,\alpha,\cdot) \tilde K(q,\alpha,q_i,\alpha_i,s,t_i) \nonumber
\end{eqnarray}
The only contribution of order $\Delta s$ is then $n=m=0$ and $l=1$ which gives 
\begin{eqnarray}
&&\int ds \int dq d\alpha  \tilde K(q_f,\alpha_f,q,\alpha,t_f,s) (-\frac{\eta}{M}) \nonumber \\  
&&\tilde K(q,\alpha,q_i,\alpha_i,s,t_i) =-\frac{(t_f-t_i)\eta}{M}  \tilde K(q_f,\alpha_f,q_i,\alpha_i,t_f,t_i)  \nonumber
\end{eqnarray}
This leads to the very simple classical contribution:
\begin{equation}
 -\frac{\eta}{M}(t_f-t_i) 
\label{eq:classical-Sb-integral}
\end{equation}
By one integration by parts (\ref{eq:classical-Si}) and (\ref{eq:classical-Sb-integral})
can be combined to
\begin{eqnarray}
&&\frac{\int\! ds dq dp  P(q,p,\cdot)
(-\frac{2\eta}{M}p\partial_{p}- \frac{\eta}{M})P(\overline{q_f},\overline{p_f},q,p,\cdot)}
{P(\overline{q_f},\overline{p_f},\overline{q_i},\overline{p_i},t_f,t_i)}
\label{eq:classical-Si-final}
\end{eqnarray}
We will show in Section~\ref{s:interpretation} that (\ref{eq:classical-Si-final}) 
together are nothing but $\overline{\beta \delta Q_{noise}}$,
where  $\delta Q_{noise}$ is the energy (heat) transferred from the system to the bath by the random force.

\subsection{The $\Delta S_b$ contribution}
\label{ss:S-contribution}
To compute the classical limit of this term we consider directly the Wigner transform $P$ of the Feynman-Vernon
propagator over the whole time interval and interpret $\frac{i}{\hbar}\alpha_i$ multiplying $\tilde K$ as $-\partial_{p_i}$
acting on $P$, and similarly $\frac{i}{\hbar}\alpha_f$ as $\partial_{p_f}$.
This gives
\begin{equation}
(-\eta)\frac{(\overline{q_i}\partial_{p_i}+\overline{q_f}\partial_{p_f})P(\overline{q_f},\overline{p_f},\overline{q_i},\overline{p_i},t_f,t_i)}
{P(\overline{q_f},\overline{p_f},\overline{q_i},\overline{p_i},t_f,t_i)}
\label{eq:classical-Sb-boundary}
\end{equation}
As (\ref{eq:classical-Sb-boundary})
is complete differential (not a proper functional), it is obviously very different from a classical entropy production term, 
and more akin to a change in state function.
In addition, it depends explicitly on initial and final position for which there is no analogy in stochastic thermodynamics.
We will return to a discussion of  (\ref{eq:classical-Sb-boundary}) in Section~\ref{s:asymptotic-analysis} below.

\subsection{The ${\cal I}^{(3)}$ contribution}
\label{ss:I3-contribution}
For the ${\cal I}^{(3)}$ part we finally have
\begin{eqnarray}
&&\int ds \int dq d\alpha dq' d\alpha' \tilde K(q_f,\alpha_f,q',\alpha',\cdot)
(\frac{\beta\eta}{(\Delta s)^2})\nonumber \\
&&\left((\Delta q)^2 -\frac{1}{4}(\Delta\alpha)^2\right) 
\tilde K_{\Delta s}(q',\alpha',q,\alpha,\cdot) \tilde K(q,\alpha,q_i,\alpha_i,\cdot) \nonumber
\end{eqnarray}
The integral over $\Delta q$ of the term in $(\Delta q)^2$ in the inner parenthesis can be evaluated as
$(\frac{i\hbar\Delta s}{M})^{n+2}\delta^{(n+2)}(\Delta\alpha-\frac{\eta}{M}\alpha'\Delta s)$
which selects  $n=0$ and $m=l=2$. The combination of pre-factors multiplying $\partial^2_{\alpha\alpha}\tilde K$
is then $\frac{\beta\eta}{M^2}(-\hbar^2)$ and as above we can interpret $-\hbar^2\partial^2_{\alpha\alpha}\tilde K$ to be $p^2$
acting on $P$. This gives a classical contribution to $\overline{\delta S_q}$ from ${\cal I}^{(3)}$ as 
\begin{equation}
\frac{\int ds \int dq dp P(q,p,\overline{q_i},\overline{p_i},s,t_i)\beta\eta(\frac{p}{M})^2
P(\overline{q_f},\overline{p_f},q,p,t_f,s)}
{P(\overline{q_f},\overline{p_f},\overline{q_i},\overline{p_i},t_f,t_i)}
\label{eq:classical-I3}
\end{equation}
As already remarked above, this quantity is $\overline{\beta Q_{friction}}$.

The remaining term $-\frac{1}{4}(\Delta \alpha)^2$ in the inner parenthesis above selects $l=n=0$ and $m=2$ giving 
$\beta\eta(-\frac{1}{2})(\frac{\eta}{M})^2\alpha^2$.
This is the same contribution as from  ${\cal I}^{(1)}$, up to the dimensionless factor $-(\frac{\beta\eta\hbar}{2M})^2$.
Since $\beta\hbar$ is the decorrelation time of the bath and $\eta/M$ is the (mesoscopic) Langevin relaxation time of the system
this factor must be very small in the set-up considered here, and can therefore be ignored.

\section{Interpretations as stochastic functionals}
\label{s:interpretation}
The purpose of this Section if to interpret all the terms derived above except 
$\Delta S_b$  
as expectation values with respect to the (classical) Kramers-Langevin process.
We begin with the term from  ${\cal I}^{(3)}$ in (\ref{eq:classical-I3}),
and express it symbolically as
\begin{equation}
\frac{\int_{q_i,p_i}^{q_f,p_f} {\cal D}(\hbox{path})\hbox{Prob}(\hbox{path})\left(\int \beta\eta (\frac{p_s}{M})^2 ds\right)}
{\int_{q_i,p_i}^{q_f,p_f} {\cal D}(\hbox{path})\hbox{Prob}(\hbox{path})}
\end{equation}
The ratio $\hbox{Prob}(\hbox{path})/\int_{q_i,p_i}^{q_f,p_f} {\cal D}(\hbox{path})\hbox{Prob}(\hbox{path})$ is the conditional probability that a given
path is chosen among the set that starts at $(q_i,p_i)$ and ends up at  $(q_f,p_f)$. Therefore we may give the interpetation
\begin{equation}
\hbox{Eq~(\ref{eq:classical-I3})}=\hbox{E}[\int \beta\eta(\frac{p_s}{M})^2 ds | q_i,p_i,q_f,p_f]
\label{eq:I3-interpretation}
\end{equation}
which is $\overline{\beta \delta Q_{friction}}$ -- as already obtained by a simpler argument above.

For the term in (\ref{eq:classical-Si-final}) we want to compare to the energy transferred to the bath from the
system by the fluctuating force. 
To the Fokker-Planck equation corresponds a Kramers-Langevin equation
\begin{equation}
\dot{p}=F -\eta\frac{p}{M} + F_{noise} \qquad \dot{q}=\frac{p}{M}
\label{eq:Kramers-Langevin}
\end{equation} 
with a random force $F_{noise}=\sqrt{2k_BT\eta}\dot{\zeta}$ where $d\zeta$ is a standard Wiener increment. To this force there is a reaction force
$-F_{noise}$ from the system on the bath and the infintessimal energy transferred to the bath is the work done by this
force,  $-\sqrt{2k_BT\eta}\dot{\zeta}\circ dq$, where $\circ$ indicates the mid-point (Stratonovich) presecription.
The average of the (classical) entropy production due to the random force is thus
\begin{equation}
\overline{\beta Q_{noise}}=\hbox{E}\left[\int -\beta\frac{p}{M}\circ \sqrt{2k_BT\eta} d\zeta  | q_i,p_i,q_f,p_f\right]
\label{eq:Q-noise}
\end{equation} 
Using alternatively $F_{noise}=\dot{p}-F +\eta\frac{p}{M}$ we should hence compute 
\begin{equation}
\int_{q_i,p_i}^{q_f,p_f} {\cal D}(\hbox{path}) \hbox{Prob}(\hbox{path})\left[ \int -\beta\frac{p}{M} \circ  \left(dp-Fds +\eta\frac{p}{M}\right)\right] \nonumber
\label{eq:Q-noise-unnormalized}
\end{equation}
We do this by discretizing the time in steps $t_i=t_0,t_1,\ldots,t_N=t_f$ and using the propagator of the Kramers-Langevin equation
\begin{eqnarray}
&& \sum_{n=1}^N\int dq_n dp_n dq_{n-1} dp_{n-1} \int -\frac{p_n+p_{n-1}}{2M}\nonumber \\ 
&&\beta\left(p_n-p_{n-1}-F_nds +\eta\frac{p_n+p_{n-1}}{2M}ds\right) \nonumber \\
&& P(q_{n-1},p_{n-1},\overline{q_i},\overline{p_i})P(q_n,p_n,q_{n-1},p_{n-1})P(\overline{q_f},\overline{p_f},q_n,p_n) \nonumber
\end{eqnarray}
The short-time propagator is
\begin{eqnarray}
P_{\Delta s}(\cdot)&=&\delta(q'-q-\frac{p'+p}{2M}\Delta s) f(\cdot) \nonumber \\
f(\cdot)&=&\frac{1}{\cal N} e^{-\frac{1}{4k_BT\eta\Delta s}\left(p'-p-F\Delta s +\eta\frac{p'+p}{2M}\Delta s\right)^2} \nonumber 
\label{eq:Kramers-Langevin-short}
\end{eqnarray}
and we therefore have 
\begin{equation}
\beta\left(p'-p-F\Delta s +\eta\frac{p'+p}{2M}\Delta s\right) P_{\Delta s}= -2\eta \Delta s \delta(\cdot) \partial_{p'}f \nonumber
\label{eq:Si-expression}
\end{equation}
Integration by parts gives three terms where the derivate 
$\partial_{p'}$ is moved respectively to $\delta(\cdot)$, $\frac{p'+p}{2M}$ or
$P(\overline{q_f},\overline{p_f},q',p')$.
The first term will be overall quadratic in $\Delta s$,
and the other two can be compared to (\ref{eq:classical-Si-final}).
Hence we have indeed that (\ref{eq:classical-Si-final})
is equal to $\overline{\beta Q_{noise}}$. Combining 
(\ref{eq:I3-interpretation}) and (\ref{eq:Si-expression}) we have $\overline{\delta S_{env}}$,
the average of the entropy production in stochastic thermodynamics,
as announced in the Introduction.

The term from ${\cal I}^{(1)}$ in (\ref{eq:classical-I1})
can also be given a probabilistic interpretation, albeit not a standard one in stochastic thermodynamics.
We start by observing that the increments of a standard Wiener process are Gaussian distributed
and that for an unconstrained average $\hbox{E}[(d\zeta)^2]=ds$ with no term of order $(ds)^2$. 
If however we average over the paths of the stochastic process that start at $(q_i,p_i)$ and end at $(q_f,p_f)$ 
we can have
\begin{eqnarray}
&&\frac{\int_{q_i,p_i}^{q_f,p_f} {\cal D}(\hbox{path})\hbox{Prob}(\hbox{path})\frac{\left(dp-Fd s +\eta\frac{p}{2M}ds\right)^2}{2k_B T\eta}}
{\int_{q_i,p_i}^{q_f,p_f} {\cal D}(\hbox{path})\hbox{Prob}(\hbox{path})} \nonumber \\
&&\qquad =ds + b[q,p,s|q_i,p_i,q_f,p_f](ds)^2
\label{eq:b-ansatz}
\end{eqnarray}
with a non-trivial coefficient $b$. We relate the expression in (\ref{eq:classical-I1}) to such a term
by observing, in analogy to (\ref{eq:Si-expression}), that 
\begin{eqnarray}
&&\frac{\left(p'-p-F\Delta s +\eta\frac{p'+p}{2M}\Delta s\right)^2}{2k_B T\eta} P_{\Delta s}= \nonumber \\
&&\delta(\cdot) \left(2k_B T\eta (\Delta s)^2 \partial^2_{p'p'}f + \Delta s f\right)
\end{eqnarray}
By two integrations by parts we therefore find that (\ref{eq:classical-I1}) is also
an average over the stochastic process
\begin{equation}
\overline{\delta S_{var}}= \int_{t_i}^{t_f}\!ds \,\frac{b}{2}[q,p,s|q_i,p_i,q_f,p_f]
\label{eq:S-var-functional}
\end{equation}
where $b$ is defined by the ansatz in (\ref{eq:b-ansatz}).
Formally we could also write (\ref{eq:S-var-functional}) as
\begin{equation}
\overline{\delta S_{var}}= \int_{t_i}^{t_f}\!ds \frac{1}{2}\,\hbox{Finite}[F_{noise}^2] \nonumber\quad\hbox{(Formal)}
\label{eq:S-var-functional-2}
\end{equation} 
where we mean the finite remainder after a term diverging as $(ds)^{-1}$ has been subtracted
from the random force squared. 

\section{Asymptotic analysis of $\Delta S_b$}
\label{s:asymptotic-analysis}
The most surprising term that have come from the above analysis is
the $\Delta S_b$ given in (\ref{eq:classical-Sb-boundary}),
both because it is like a change in a state
function, and also because of its dependence on initial and final position.
We will here consider this term in the limit of weak coupling.
As this is essentially a classical problem we will adopt the
dimension-less units introduced below in 
Appendix~\ref{s:times} where the Kramers-Langevin equation reads
\begin{equation}
\dot{p}=f -\gamma p + \sqrt{2\gamma}\dot{\zeta}  \qquad \dot{q}=p
\label{eq:Kramers-Langevin-nondim}
\end{equation} 
where $\zeta$ is standard white noise and where
\begin{equation}
\Delta S_b = -\gamma\left(q_i\partial_{p_i}+q_f\partial_{p_f}\right)\log P 
\label{eq:Sb-nondim}
\end{equation} 
where $P$ is the transition probability of (\ref{eq:Kramers-Langevin-nondim}).
We will only consider the case when $t$, the duration of the process, is of order one
in (\ref{eq:Kramers-Langevin-nondim}) ($t$ on the order of $t_{osc}$, the characteristic time of the
system, in the original dimensional variables), and $\gamma$ tending to zero.
After the limit in $\gamma$ is taken we could also allow $t$ to become long.
We note that the opposite limit where first $t$ is taken long and then $\gamma$ is taken to 
zero is physically more interesting, but also mathematically considerably more 
complicated~\cite{FreidlinWentzell04,LamKurchan2014}, and outside the scope of this discussion.

Let ($q_f^*,p_f^*$) be the final position and momentum at time $t$ of the classical conservative system
defined by (\ref{eq:Kramers-Langevin-nondim}) when $\gamma=0$, starting from ($q_i,p_i$).
If $\Delta q_f=q_f-q_f^*$ and $\Delta p_f=p_f-p_f^*$ are the deviations of the actual
final positions from the classical path we assume $P(q_f,p_f,q_i,p_i)$ to be a Gaussian distribution, \textit{i.e.}
\begin{equation}
P(\cdot)\sim e^{-\frac{1}{\gamma t}\Sigma}
\label{eq:LD}
\end{equation}
where 
\begin{equation}
\Sigma = \frac{1}{2}(\Delta p_f,\Delta q_f)C^{-1} (\Delta p_f,\Delta q_f)^T
\end{equation}
and where $C$ is the
correlation matrix of the deviations in units of $\gamma t$.
The argument for this is linearizing (\ref{eq:Kramers-Langevin-nondim}) to
\begin{equation}
\dot{\Delta p}=f'(q)\Delta q -\gamma \Delta p + \sqrt{2\gamma}\dot{\zeta}  \qquad \dot{\Delta q}=\Delta p
\label{eq:Kramers-Langevin-nondim-linearized}
\end{equation} 
and noting that if we ignore the terms $f'(q)\Delta q$ and $\gamma \Delta p$, the distribution of
of ($\Delta p, \Delta q$) is that of Brownian motion and integral of Brownian motion.
These two are jointly Gaussian distributed, with (note that $\gamma t$ is explicitly included in (\ref{eq:LD}))
\begin{equation}
\Sigma = 2 \left(\begin{array}{ll} 1 & t/2\\
                                   t/2 & t^2/3
            \end{array}\right) \hbox{(Brownian motion)} \nonumber
\end{equation}
Under these assumptions we have the two contributions to $\Delta S_b$ as
\begin{eqnarray}
\frac{1}{t} q_f \partial_{p_f}\Sigma &=& \frac{1}{t} q_f  \left(C^{-1}_{11}\Delta p_f + C^{-1}_{12}\Delta q_f\right) \\
\frac{1}{t} q_i \partial_{p_i}\Sigma &=& \frac{1}{t} q_i  (-\frac{\partial p_f^*}{\partial p_i},-\frac{\partial q_f^*}{\partial p_i})C^{-1} (\Delta p_f,\Delta q_f)^T 
\label{eq:Sb-nondim}
\end{eqnarray} 
These terms do not go to zero with $\gamma$. On the other hand, ($\Delta p_f,\Delta q_f$) is
by assumption of typically of order $\sqrt{\gamma t}$, and the probability that
$\Delta S_b$ is of order one hence exponentially small in $\gamma^{-1}$.

\section{Discussion}
\label{s:discussion}
In this contribution we have a computed a entropy production in a bath coupled
to a driven quantum system defined as the  
change of the von Neumann entropy of the bath.
Except for the (central) assumption that the bath is comprised of 
a set of harmonic oscillators initially in thermal
equilibrium and interacting linearly with the system,
the analysis is general. A main result is 
that this entropy production can be written as a (quantum) expectation
value of three functionals given in equations (\ref{eq:I-definition}) and (\ref{eq:h-definition}).

The classical limit of these functionals is however puzzling. While we do recover 
$\overline{\delta S_{env}}$, the average over a classical Kramers-Langevin process
and for a finite time of the standard entropy functional $\delta S_{env}$,
we also find another functional which we have here called $\overline{\delta S_{var}}$,
and a boundary contribution $\Delta S_b$. The latter is hence not a proper
functional but more akin to change in the internal state of the system.
As we have shown above, in the weak coupling limit $\Delta S_b$ does not
vanish except when the initial and final points are on the same
classical path, an unsatisfactory result from a physical point of view.
A possible way out is that in this limit, and the assumptions made, the system
is (almost) conservative and follows (nearly) the classical equations of motion. 
Deviations from the classical path leading to non-zero $\Delta S_b$ are very rare
and mean that 
the random force $F_{noise}$ from the bath has a considerable effect on the system, 
even if this force has a very small amplitude. The degrees of freedom of the bath 
must then exercise some kind of coordinated action which is untypical of a thermal state, and the 
starting assumption made in Eq.~(\ref{eq:dTrLogRho}), that it suffices to consider a first-order
change of the density operator of the bath away from thermal equilibrium, is at least questionable.
More work will be needed to find out if this is a viable explanation.

\section*{Acknowledgements}
E.A. thanks Karol \.{Z}yczkowski, Jakub Zakrzewski and the participants
of the conference ``AV60'', Rome, Italy (September 2014) for valuable discussions,
and the Kavli Institute for Theoretical Physics China (October 2014) for hospitality. 
%Eq~(\ref{eq:I-definition}) above was independently derived by Ralf Eichhorn. 
This research is supported by the Swedish Science Council through grant 621-2012-2982,
and by the Academy of Finland through its Center of Excellence COIN.

\appendix

\section{Time scales}
\label{s:times}
In this Appendix we discuss for completeness what means the 
assumption that $\hbar\beta$ is much smaller than
all times scales of interest in the system.
Our starting point is that the
short-time expression of the Feynman-Vernon propagator (\ref{eq:short-time})
should then be valid for $\Delta s=\hbar\beta$. 
This is so for any constant potential $V$
hence a restriction from the system can only come from spatial variations of $V$.
We use that the increments that matter are those for which
$(\Delta X)^2\sim\frac{\hbar^2\beta}{M}$ and
$(\Delta Y)^2\sim\frac{\hbar^2\beta}{M}$.
and assume that $V=E U(q/\ell)$
where $E$ is a characteristic energy scale, $\ell$ is a characteristic length, and $U$ is a dimension-less function of order 
unity.
The variation of the potential term in the action -- with respect to constant potential -- over a time $\hbar\beta$ is then
\begin{equation}
\delta\left(\frac{i}{\hbar}\int_s^{s+\hbar\beta}V(q_s')ds'\right) \approx (\hbar\beta)\cdot \frac{E}{\ell\hbar}\cdot \left(\frac{\hbar^2\beta}{M}\right)^{\frac{1}{2}}
\label{eq:balance}
\end{equation}
The condition that (\ref{eq:balance}) is much less than one means 
\begin{equation}
\hbar\beta << (\beta E)^{-\frac{1}{2}} \cdot \sqrt{\frac{M}{E}}\ell
\end{equation}
The second term on the right-hand side is a characteristic time of the system $t_{osc}$ which can alternatively
be found as the period of an oscillation, assuming that the potential somewhere has a quadratic minimum such
that $V\sim \frac{1}{2}\frac{E}{\ell^2} (q-q^*)^2$.
The first term on the right-hand side is dimension-less, and by the standard assumption of
stochastic thermodynamics that thermal and mechanic energy are comparable it can be taken of order unity.

We hence have three characteristic times: $\hbar\beta$, which is a property of the bath,
$t_{osc}$, which is a property of the system, and the Langevin relaxation time $M/\eta$,
which is a property of the interaction between the system and the bath,
and by assumption $\hbar\beta<< t_{osc}$ and $\hbar\beta << M/\eta$.
Introducing the dimension-less ratio $\gamma=\frac{\eta t_{osc}}{M}$,
the new dimension-less variables $t\to \frac{t-t_i}{t_{osc}}$, $q\to q/\ell$, $p\to pt_{osc}/M\ell$,
and writing the force $F=\frac{1}{\beta \ell}f(\xi)$ where $f(\xi)=-U'(\xi)$,
we have the Kramers-Langevin equation as given in (\ref{eq:Kramers-Langevin-nondim})
in the main text.

\section{Higher-order corrections to the Caldeira-Leggett theory}
\label{s:higher}
In this appendix we compare the the first correction to the Fokker-Planck equation (\ref{eq:CL-FokkerPlanckeq})
which stems from the third-order derivative of the potential, to a higher-order correction
to the Caldeira-Leggett theory. It is easily seen that the third-order term is
$-\frac{\hbar^2}{24}\partial_{ppp}(V^{(3)}P)$. The Wigner function would hence to this order
obey a partial differential equation, albeit not a Fokker-Planck equation, and we will see that this is true
for the higher-order correction to the Caldeira-Leggett theory as well.

The starting point in computing the higher-order correction is to expand the $\coth$ function in the
Feynman-Vernon action $S_r$ to next order \textit{i.e.} $\coth(\frac{\beta\hbar\omega}{2})= \frac{2}{\beta\hbar\omega} + \frac{1}{6}\beta\hbar\omega+\ldots$.
This gives a first-order correction $\sum_b \frac{C_b^2\beta\hbar}{12m_b}\cos\omega_b(s-s')$ to the kernel $k_r$. 
A term in this kernel independent of frequency will in the limit tend to the second derivative of a delta function
(as we have seen above for ${\cal I}^{(3)}$) which hence gives a first-order correction $-\frac{\beta\eta\hbar}{6}\int \dot{\alpha}^2 ds$
to $S_r$. This is a kinetic term, which hence gives a correction to the Feynman-Vernon short-time propagator
$e^{-\frac{\beta\eta\hbar}{6\Delta s}(\Delta\alpha)^2}$, smaller than the main kinetic term by the
dimension-less ratio $\frac{\beta\hbar}{M/\eta}$. We assume this ratio small and expand the correction
to the kinetic term as $1-\frac{\beta\eta\hbar}{6\Delta s}(\Delta\alpha)^2+\ldots$ and carry out the integration 
over $\Delta q$ as in (\ref{eq:expansion}). The presence of an additional term $(\Delta\alpha)^2$, which can
be expanded as $(\Delta\alpha-\frac{\eta}{M}\alpha'\Delta s)^2+\ldots$ allows for a second-order term $(\Delta q)^2$ in
the expansion in  (\ref{eq:expansion}), which then results in an additional term
$\frac{\hbar}{6M}\frac{\beta\eta\hbar}{M}\partial^2_{qq}\tilde K$ in
(\ref{eq:CL-Lindbladeq}). 
That leads a Wigner transform obeying 
(\ref{eq:CL-FokkerPlanckeq}) plus a term
$\frac{\hbar}{6M}\frac{\beta\eta\hbar}{M}\partial^2_{qq}P$
which has the meaning of a diffusion acting directly on position.

To differentiate the two terms derived in this section,
which are both quadratic in $\hbar$, we use the dimension-less variables introduced in Appendix~\ref{s:times} above
and write the generalization of (\ref{eq:CL-FokkerPlanckeq}) as 
\begin{eqnarray}
\partial_{t} P&=& -\partial_{q}(p P)-\partial_{p}\left((f-\gamma p)P\right)+
\gamma\partial_{pp}P \nonumber +\\
&&(\frac{\hbar\beta}{t_{osc}})^2\left(\frac{1}{24}\partial_{ppp}(f^{(2)}P)
+\frac{\gamma}{6}\partial_{qq}P\right)+\hbox{h.o.t} \nonumber
\end{eqnarray}
where $\gamma=\eta t_{osc}/M$ is the ratio of 
the characteristic time of the system and the Langevin relaxation time.
In principle the higher-order correction to the Caldeira-Leggett theory is hence more important
than expanding the potential $V$ if either $\gamma$ is much larger than unity, or if $f^{(2)}\approx 0$.
The first case however corresponds to the overdamped limit classically described 
by spatial diffusion with diffusion coefficient $\gamma^{-1}$, which is much
larger than the spatial diffusion term derived here unless
$M/\eta \sim \hbar\beta$, while one of the assumptions made above was that $M/\eta >> \hbar\beta$.
Therefore it is only consistent to retain only the higher-order correction 
if the potential is approximately harmonic, as then in fact $f^{(2)}\approx 0$.
\bibliography{fluctuations,AZZNotes}%
\end{document}